\documentstyle[preprint,aps]{revtex}  
\title{Remark on the Second Principle of Thermodynamics}
\author{Constantino Tsallis}
\address{Centro Brasileiro de Pesquisas Fisicas,
 Rua Xavier Sigaud 150, 22290-180 Rio de Janeiro-RJ, Brazil  (tsallis@cbpf.br)}

\draft

\begin{document}

\maketitle

\begin{abstract}

All presently available results lead to the conclusion that nonextensivity, in the sense of nonextensive statistical mechanics (i.e., $q \ne 1$), does {\it not} modify anything to the second principle of thermodynamics, which therefore holds in the usual way. Moreover, some claims in the literature that this principle can be violated for specific anomalous systems (e.g., granular materials) can be shown to be fallacious. One recent such example is analyzed, and it is suggested how $q \ne 1$ distributions could in fact restore the validity of macroscopic time irreversibility, a cornerstone of our present understanding of nature.

\end{abstract}

\bigskip

The Second Principle of Thermodynamics, i.e., time irreversibility of the macroscopic world, appears to be one of the most solid laws of theoretical physics. Usual statistical mechanics, i.e., Boltzmann-Gibbs statistical mechanics, is consistent with this principle through the celebrated $H$-theorem. We address here what happens with this principle within the nonextensive statistical mechanics proposed in 1988\cite{tsallis} and characterized by an entropic index $q$ (the particular case $q=1$ recovers standard Boltzmann-Gibbs statistical mechanics). The answer seems to be very simple: no hope for {\it moto perpetuo} within this formalism! Indeed, the $H$-theorem appears to be $q$-invariant, more precisely, excepting for very quick microscopic fluctuations, $S_q$ cannot decrease (increase) with time if $q>0$ ($q<0$), a fact which is consistent with the concavity (convexity) of $S_q$, with regard to the set of probabilities, for $q>0$ ($q<0$) (hence, $S_q$ is respectively maximal and minimal for $q>0$ and $q<0$). This $q$-invariance was first pointed by Mariz\cite{mariz} assuming detailed balance, and since then by many other authors (see \cite{various} and references therein) in a variety of generic situations. In addition to this, the present nonextensive thermostatistics provides a possible path for solving amazing "violations" of the second principle of thermodynamics. The rest of this paper is dedicated to the re-analysis of one such example that appeared recently in the literature of granular materials. 

Through molecular dynamics simulations using the inelastic hard sphere model to mimic fluidized granular media, Soto, Mareschal and Risso addressed \cite{soto} a very interesting question, namely the validity (or not) of Fourier's law and of the second principle of thermodynamics for dissipative macroscopic systems. They simulated a simple $d=2$ system in the presence of a "vertical" ($y$ axis) gravitational field, the top and bottom edges of a square box providing, to the arriving particles, a Maxwellian distribution of velocities at a fixed (dimensionless) temperature $T=1$. The average density $n_0$ and  the dissipation coefficient $q_{SMR}$ (we use here the notation $q_{SMR}$ instead of the original notation $q$\cite{soto} in order to avoid confusion with the nonextensive entropic index $q$) were held fixed during a given simulation ($q_{SMR}=0$ corresponds to the conservative limit, i.e., elastic collisions). They measured, as functions of the height $y$, the density $n$, the temperature $T$ and the heat flux ${\bf J}$. After long runs the system achieved a stationary state, like the one illustrated in their Fig. 1. The authors verified that $T(y)$ exhibits a minimum at $y=y^*$, whereas ${\bf J}$ vanishes at a value of $y$ {\it larger} than $y^*$. Consequently, the authors conclude that Fourier's law, which reads ${\bf J}=-k \nabla T$, $k$ being the thermal conductivity, is violated. Moreover, for all values of $y$ not exceedingly above $y^*$, the heat appears to flow from the cold regions to the hot regions, therefore the authors consistently conclude that the second law of thermodynamics is violated as well. The way out from this paradoxal situation that Soto et al advocate is that an extra, $n$-dependent, term must be included in the expression of ${\bf J}$.

The purpose of the present remark is to argue that the proof provided by Soto et al would be very neat were it not a subtle weakness that we now address. The quantities $n$ and ${\bf J}$ observed in the simulations are calculated essentially through their definitions. This is {\it not} the case of $T$, which the authors set {\it equal to the average kinetic energy per particle}. This equality holds for the well known Maxwellian distribution, {\it an assumption which is certainly false inside the box} (see, for instance, \cite{taguchi,kadanoff}). In order to illustrate the subtleties, let us use 
nonextensive statistical mechanics. Indeed, the use of such generalized statistics for the present system is quite natural since the stationary distribution observed by Taguchi and Takayasu in a vibrating-bed fluidized state {\it precisely} corresponds, as we shall see, to $q=3$ (similarly to what occurs in L\'evy  and  correlated anomalous diffusions \cite{levy}), whereas they observe a Maxwellian distribution (i.e., $q=1$) in the so-called solid state. The possibility of applicability of this formalism to such materials has also been recently mentioned by Herrmann\cite{herrmann}. 

For the Hamiltonian ${\cal H}=   \frac{1}{2}  \sum_{i=1}^N {\bf v}_i^2$, the nonextensive formalism yields, for the velocity distribution probability,
\begin{eqnarray}
p_q(\{ {\bf v}_i \})=  
\frac{\Bigl[1-(1-q)  ({\cal H}-U_q) /T_q \Bigr]^{\frac{1}{1-q}} }{Z_q}
\end{eqnarray}
where $Z_q \equiv  
\int \Bigl[ \Pi_{j=1}^N d{\bf v}_j  \Bigr] 
   \Bigl[1-(1-q)  ( {\cal H}-U_q) /T_q \Bigr]^{\frac{1}{1-q}}$, $U_q \equiv \int \Bigl[ \Pi_{j=1}^N  d{\bf v}_j \Bigr] p_q^q \;  {\cal H} / \int \Bigl[ \Pi_{j=1}^N d{\bf v}_j\Bigr] p_q^q$ and $T_q \equiv T Z_q^{1-q}$.
This equilibrium distribution has been obtained by optimizing 
$S_q = \Bigl(1-\int \Bigl[ \Pi_{j=1}^N d{\bf v}_j\Bigr] p^q \Bigr) / (q-1)$ with appropriate constraints (Boltzmann constant $k_B$ has been taken equal to unity). It recovers the usual Maxwellian distribution for $q=1$ and a power-law for $q \ne 1$. Also, we verify that $1/T=\partial S_q/\partial U_q$. Finally, we can rewrite Eq. (1) as follows:
\begin{eqnarray}
p_q= 
\frac{\Bigl[1-(1-q)  {\cal H}  /T^{\prime} \Bigr]^{\frac{1}{1-q}} }{Z_q^{\prime}}
\end{eqnarray}
where $Z_q^{\prime} \equiv  
\int \Bigl[ \Pi_{j=1}^N d{\bf v}_j\Bigr] 
   \Bigl[1-(1-q)   {\cal H} /T^{\prime} \Bigr]^{\frac{1}{1-q}}$ 
and $T^{\prime} \equiv T_q+(1-q)U_q$.

We verify that, for $q=1$, $T_q=T^{\prime}=T$, but such simplification disappears if $q \ne 1$, which might be the case for the system considered by Soto et al. In particular, if we identify $p_q^q$ ($p_q$ being the $N=1$ distribution (2)) with the distribution numerically obtained in \cite{taguchi} for the vibrating-bed fluidized state, we obtain $q/(q-1)=3/2$, hence $q=3$.  In other words, some further considerations are necessary before drawing any conclusion about the validity (or not) of Fourier's law and of the second principle of thermodynamics. A remarkable simplification does exist due to the fact that Yamano and others (see \cite{others} and references therein) have recently shown \cite{yamano} that each quadratic velocity term contributes to $U_q$ with $T_q/2$. But, for the rest, the entire Fourier law would have to be rededuced, possibly in the $q \ne 1$ scenario. A variety of mathematical tools \cite{rajagopal} (e.g., the $q$-generalization of Kubo's linear response theory) are already available in the literature for performing such generalization. It is clear nevertheless that this is not a trivial task. Only {\it after} such an analysis, the interesting computer simulations of Soto et al could led us to further conclusions. It could even happen that, in the present granular matter fluidized state, the generalized ${\bf J}$ vanishes precisely when $T_q$ attains its minimum!

\end{document}